\documentclass{sig-alternate-05-2015}
\usepackage{color}
\usepackage[square,numbers,sort&compress]{natbib}
\usepackage{eurosym}
\usepackage{gensymb} 
\usepackage{paralist}
\usepackage{nameref}

\newcommand\blfootnote[1]{%
  \begingroup
  \renewcommand\thefootnote{}\footnote{#1}%
  \addtocounter{footnote}{-1}%
  \endgroup
}

\newcommand{\subparagraph}{}
\usepackage[compact]{titlesec}

\usepackage{todonotes}

\begin{document}

\setcopyright{acmcopyright}



\conferenceinfo{HIP3ES'17,}{ January 25, 2017, Stockholm, Sweden}


%
\CopyrightYear{2017} 

\title{Power and Execution Time Measurement Methodology for SDF 
Applications on FPGA-based MPSoCs}
%
%
%
%
%

\numberofauthors{3} 
%
\author{
%
%
\alignauthor
Christof Schlaak\\
       \affaddr{OFFIS Institute for\\ Information Technology}\\
       \affaddr{Oldenburg, Germany}\\
       \email{christof.schlaak@offis.de}
\alignauthor
Maher Fakih\\
       \affaddr{OFFIS Institute for\\ Information Technology}\\
       \affaddr{Oldenburg, Germany}\\
       \email{maher.fakih@offis.de}
\alignauthor
Ralf Stemmer\\
       \affaddr{University of Oldenburg, Germany}\\
       \email{ralf.stemmer@uni-oldenburg.de}
}
\date{15 September 2016}

\maketitle

\begin{abstract}
Timing and power consumption play an important role in the design of embedded systems. Furthermore, both properties are directly related to the safety requirements of many embedded systems.
With regard to availability requirements, power considerations are of uttermost importance for battery-operated systems. 
Validation of timing and power requires observability of these properties. In many cases this is difficult, because the observability is either not possible or requires big extra effort in the system validation process. In this paper, we present a measurement-based approach for the joint timing and power analysis of Synchronous Dataflow (SDF) applications running on a shared memory multiprocessor systems-on-chip (MPSoC) architecture. As a proof-of-concept, we implement an MPSoC system with configurable power and timing measurement interfaces inside a Field Programmable Gate Array (FPGA). Our experiments demonstrate the viability of our approach being able of accurately analyzing different mappings of image processing applications (Sobel filter and JPEG encoder) on an FPGA-based MPSoC implementation.
\end{abstract}

%

\ccsdesc[500]{Computer systems organization~System on a chip}
\ccsdesc[500]{Computer systems organization~Embedded hardware}

%
%

%
%
\printccsdesc


\section{Introduction}
\blfootnote{\scriptsize{
This work has been partially supported by the ARAMIS II project (01|S16025J), which is funded by the German Federal Ministry of Research and Education (BMBF),
and by the SAFEPOWER project with funding from the European Union's Horizon 2020 research and innovation programme under grant agreement No 646531.
}}
Low power consumption and meeting real-time requirements are key issues in embedded systems design. 
With the growing computational demand of nowadays applications in the automotive, avionics and multimedia domain, the size and complexity of embedded systems based on multiprocessor platforms (MPSoCs) is increasing and thus leading to high power consumption. 
MPSoCs are used ubiquitously in modern designs and their power consumption may have a major impact on the overall system power consumption. Especially for mobile battery powered computers the main fraction of the overall power consumption is due to complex processing elements (radio, main processor, graphics accelerator) \cite{SmartPhonePower}.

There are mainly two approaches for the estimation of power consumption of MPSoCs: Analytical and empirical methods.
By the analytical methods a mathematical model of the device under test (DUT) is constructed where the components influencing the DUT power are modeled (e.g. circuit switching activity) and an analysis is made to obtain estimates for average/peak power consumption.
In empirical methods, power consumption is measured directly on the hardware for single devices e.g. for the processor, or for the whole MPSoC.

Power estimation of an MPSoC is not an easy task due to the lack of observability. In many cases, power measurement can only be performed at the MPSoC's power rail inputs and no direct relationship between the running software and measured power consumption can be established.
The measured power consumption consists of a static and a dynamic part. The \textit{static power contribution} depends on parameters that are fixed at MPSoC design time (chip area, used technology and process variation/corner) and dynamic properties that can be externally controlled (supply voltage, ambient temperature). In this paper, we assume static power to be constant. Although this is an oversimplification, the control of static power consumption is out of the scope of this paper.
The \textit{dynamic power contribution} (i.e. switching activity) is completely application and data dependent \cite{lee_cycle-accurate_2005} and is affected by many factors. E.g., the software functionality, mapping of software tasks to processors, software scheduling, communication between tasks and the resulting communication and computation resource utilization. In this paper, we will focus on the measurement of the dynamic power consumption.

In this paper, we present a measurement-based approach for time and power analysis of multiple Synchronous Dataflow (SDF) applications running on an MPSoC implemented on an FPGA. 
As application model we use the SDF model of computation (MoC) \cite{lee1987}, because it offers a strict separation of communication and computation. An SDF application consists of computation kernels called actors (see top of Fig.~\ref{fig_inputs}), and communication channels following the FIFO-concept.
The execution of an actor has three phases: a) The read phase where all data are read from all incoming channels; b) The computation phase, where data is processed; c) The write phase, where the actor writes the output data into the FIFO-buffers of all outgoing channels.
This allows us to analyze the communication and computation time and power consumption of our application separately. 
Our approach comes with the following contributions:

\begin{compactenum}
\item We present a measurement concept which allows a minimal invasive timing and power analysis of the SDF applications at different granularity levels e.g. for single phases (\texttt{Write, Compute, Read}) of an actor, for single SDFGs and for the whole system application,
 
\item We integrate a low-cost measurement board (\textit{Mageec} \cite{Mageec}) with a customized measurement controller (implemented in the FPGA) to realize a low-cost implementation of our flexible measurement concept, allowing power and time analysis of a given implementation in an automated way.
\end{compactenum}

\section{Related Work}
\label{sec:related}

Various approaches utilize measurement-based methods for the direct evaluation of FPGA power consumption or the validation of power models (estimations).

Schreiner et al. \cite{SGRN15} used an FPGA board with an integrated measuring electronics for this purpose. It captures the power consumption but with a low sampling rate of $ 6 {,}  25 Hz$  which is too small to enable a detailed analysis of short sub-phases of the software to be measured.

A high-level power model for FPGA-based MPSoCs was presented in \cite{piscitelli_high-level_2011}. For the evaluation of this model, the power consumption is measured and compared with the predicted values. The measurement was made with the help of the Virtex-6 FPGA with its integrated electronics that stores the measured power values in internal measurement registers (with a sampling rate of 5 Khz). 

The work in \cite{hosseinabady_run-time_2014,hosseinabady_energy_2015} exploits the clean semantics of SDF applications to apply power-gating to reduce their energy consumption when running on FPGAs. They use the on-board measurement devices of the Xilinx ZC702 board for measuring the power consumption and evaluating the efficiency of their approach. 

Also the work in \cite{beldachi_accurate_2014} suggests an accurate power consumption measurement utilizing the on-board power monitors which can be found in some FPGAs for e.g. the ZC702 Xilinx board. 
Typically, these on-board power monitors have a low sampling rate to perform detailed analysis of software applications. For e.g. one of the most modern Xilinx FPGA boards ZC702 \cite{Xil14} samples the power rails with the help of the integrated power controller Texas Instruments UCD9248 \cite {Ti12} every 200 $\mu$s (i.e., at a frequency of $5 kHz$). These sampling rates are too low to measure the power consumption of short phases of software execution on the MPSoC.

In \cite{ou_rapid_2006} a laboratory power supply with built-in measuring device (\textit{Keithley SourceMeter 2400}) is used. In \cite{SDFSEB15,oliver_power_2011, albicocco_direct_2012, JC11} oscilloscopes connected to shunt resistors are used to measure power consumption. Using oscilloscopes or specialized power measurement device can obtain accurate results with high resolutions but having the disadvantage of high costs.
 
Schabbir et. al \cite{shabbir_ca-mpsoc:_2010} presented a design flow to generate multiprocessor platforms for multiple SDFGs. In this flow, models for performance prediction are used to obtain rough estimates of the periods of the SDF implementations. To evaluate these predictions, the MPSoC is implemented on an FPGA and a set of SDFGs are executed. A hardware timer in the FPGA measures the periods of the SDF applications, so that the prediction of the measurement can be compared. This approach has many similarities with the measurement concept of this work. However, it only addresses the measurement of time; a combined power measurement is not considered. In addition, the measurement of the execution time refers only to periods, rather than more fine grained levels (e.g. measuring the delay of the actor phases).

The work in \cite{GOBHLB11} dealt with the effects of parameters such as the number of processors and the clock frequency on the performance and the power consumption of FPGA-based MPSoCs. For the characterization of the various system designs, a timer IP that is connected via a shared bus with the soft processors (MicroBlazes) captured the execution time. The power consumption is not measured but estimated via Xilinx Power Estimator (XPE) \cite{xpe} tool. The impact of the software is not covered and a detailed analysis of an SDF application (for example, the actor phases) is not possible. 

In \cite{lee_cycle-accurate_2005} a cycle-by-cycle energy measurement in FPGAs based on switched capacitor is presented. With the help of this approach, a high resolution of the measured energy values (every 20 ns) was achieved. Another measurement approach presented in \cite{Weiland2015} also achieves a high resolution and is capable of measuring SW applications with detailed granularities when running on FPGAs. 
In difference to the work above, our approach introduces a hardware component on the FPGA that flexibly trigger the power/time measurement of the annotated running application modeled as SDFGs.
Nevertheless, it is possible to use the measurement infrastructure from \cite{Weiland2015} combined with our measurement concept to analyze SDF applications.

To the best of our knowledge, no similar measurement approach was found which enables measuring the execution time and power consumption of multiple SDF applications at different granularity levels running on an FPGA-based MPSoC. Especially the usage of a low-cost measurement board (in our case the Mageec-board with hardware costs around 50\euro) for measuring the power consumption of an FPGA-based MPSoCs is novel.

\section{Measurement Concept}

\begin{figure}
\centering
\includegraphics[width=0.8\linewidth]{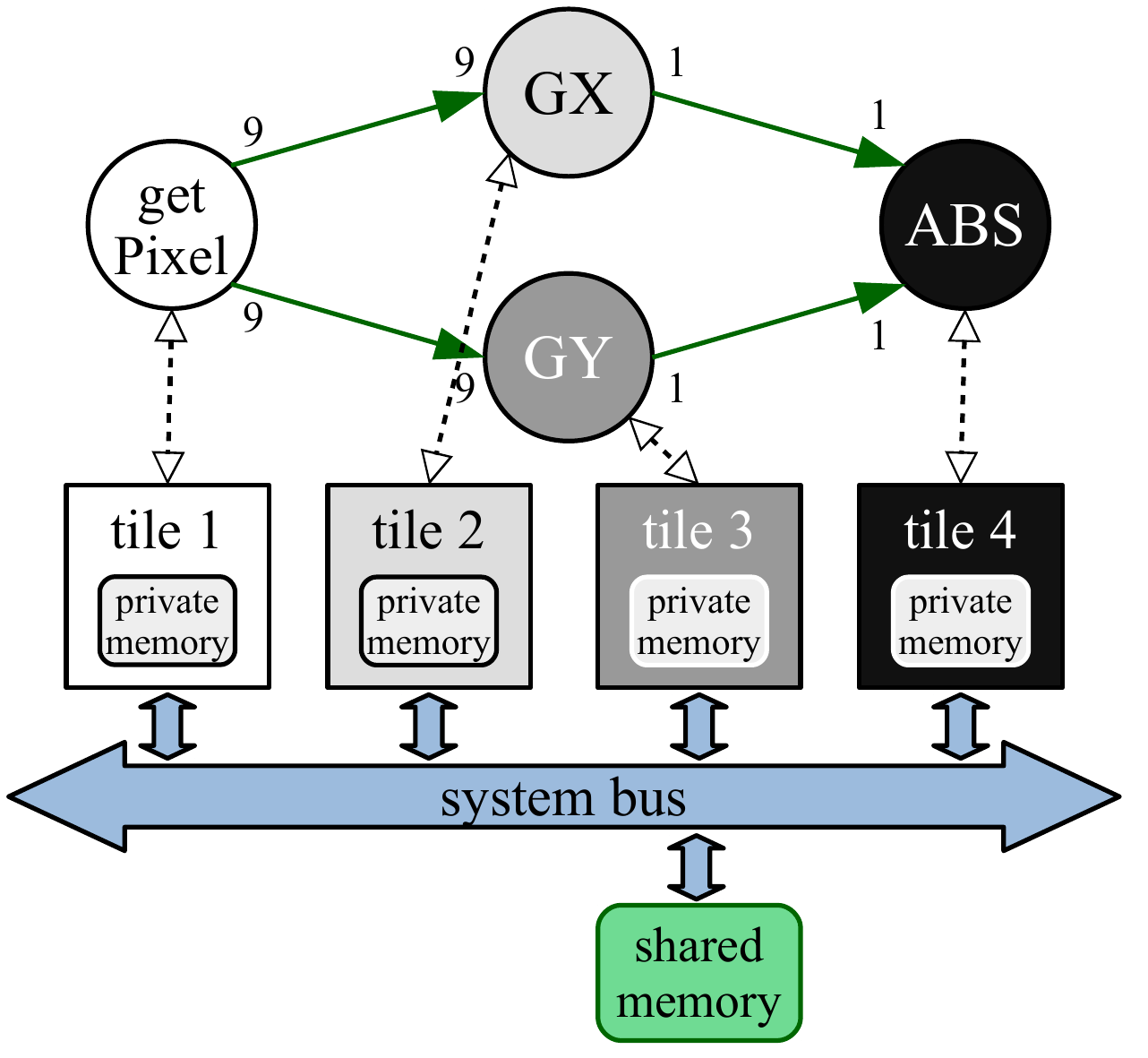}
\caption{User-specified example inputs for the analysis: A Sobel filter modeled as a SDFG (\textit{top}) representing the software application. An MPSoC (\textit{bottom}) with four processors and a shared memory connected through a shared system bus. Each processor has its own private memory. A mapping of the actors on the processors (\textit{matching colors} and \textit{dotted lines}) and of the channels on the shared memory (\textit{matching color green})}.
\label{fig_inputs}
\end{figure}

When analyzing an SDF application in detail, the level of \emph{measurement granularity} is important. Timing and power analysis of both sub-phases of an application behavior, as well as its overall behavior is relevant. We define four granularity levels (from coarse to fine) that should be supported by our approach: The \textit{System-level granularity} which is independent of time. At this level, all SDFGs are repeatedly executed, while the (average) power is measured over a period of time. On the \textit{SDFG-level granularity}, all SDFGs of the application are analyzed one after another, giving information about the timing (throughput or end-to-end latency) and power usage of every SDFG\footnote{In terms of SDFGs, an \textit{iteration} is completed when the initial tokens distribution on all its channels is restored. Having the iteration in mind, measurements can be guided to trace the activations of the actors leading to an iteration of the SDFG.}. Next, the actors of each SDFG can be analyzed at the \textit{actor-level granularity}. Here, the analysis results are useful to analyze optimal actors to tile mapping. Furthermore, the \textit{phase-level granularity} provides the most detailed analysis for every read, compute and write phase of every actor. Again, these measurements help optimizing the actor to tile mapping, considering the actor's computation and communication demands. In order to support the analysis of SDF applications at the above-defined levels, our timing and power analysis measurement approach requires the following inputs (see also Fig. \ref{fig_inputs}):

\begin{compactitem}
\item a specified \emph{granularity} level, in which the application will be explored,
\item an \emph{MPSoC} as a hardware platform implementation,
\item one or more \emph{SDFGs} implemented as a software application,
\item and a \emph{Mapping} of the SDFG parts to the MPSoC resources.
\end{compactitem}

Our measurement approach analyzes the DUT according to the chosen granularity and assigns timing and power values to each of the identified code blocks (actor phases, whole actors, iterations).
For this, each block is measured separately and several times in a \emph{measurement scenario}. By repeating each measurement, we can trace best, worst and average timing and power results. Obviously, the measurement of best- and worst-case values do not necessarily provide any guarantee on lower and upper bounds. Extensions towards a full measurement data collection (histogram) and calculation of the measurement variance are possible, but not in the scope of this work. 

\subsection{Controlling the measurement}

In order to tag the relevant phase of the application (e.g. the \texttt{Write} phase of an actor), the corresponding application code should be instrumented with measurement control signals. Before executing the relevant part of the application code, the processors of the MPSoC send \texttt{start} signals to a customized \emph{measurement controller} (see Fig.~\ref{fig_structure}). When the execution is finished, they send \texttt{stop} signals. Thus, the measurement controller is able to recognize the relevant phase and to trigger the measurement process at the right time.

In general, the aim is to keep the invasiveness caused by the code instrumentation of our approach as little as possible. This can be done by keeping the delay time of the instructions, needed to be executed on a processor to trigger the measurement control unit, minimal. Another goal is to make the communication between processor and measurement controller as fast as possible to immediately start and stop the (power) measurement and at the same time to avoid contention between concurrently accessing processors when triggering the measurement controller.

In order to achieve above goals, we instrument the original source code of the SDF application with \texttt{start} and \texttt{stop} statements that control the measurement with the help of a minimal set of instructions (where each control statement costs merely 2 cycles of delay on a MicroBlaze processor). Depending on the current scenario, the placement of these statements in the source code varies (see Sect.~\nameref{subsec:Instrumentation}). Of course, the execution time of the annotated code, when run on a target processor, is now delayed in different ways for each scenario compared to the unmodified original application code. Consequently, the application's timing behavior is not the same for the measurements and the real use-case, which is undesirable for a measurement approach. 
We coped with these timing variations by creating delay statements (consisting of NOPs (No Operation)), that take the same amount of time as the measurement control statements, when executed. After every measurement, these delay statements replace the measurement control statements automatically, enforcing the same timing behavior in the target application as the annotated one in the measurement scenarios. 
This way the annotation affects the application equally during measurement and in the real use case. Certainly, by doing this, the timing behavior of the original application has been changed. However, as we will in Sect.~\nameref{sec:Evaluation} these changes are of little account.

There are many ways to notify the measurement controller about the beginning or ending code blocks to be measured. In our measurement concept, we suggest that the processors use their peripheral buses to send triggers to the measurement control unit. On an FPGA based hardware platform, each processor can be configured such that it has an exclusive access to its own peripheral bus to avoid contention caused by concurrent accesses of multiple processors. When using COTS (Component-Of-The-Shelf) MPSoC as target platform, we do not necessarily have the flexibility reserve a peripheral bus uniquely for every processor. In that case, the worst-case delay, which is raised by simultaneous bus accesses of the participants (e.g. processors), must be taken into account. In this paper, we focused on dealing with FPGA based MPSoCs. Nevertheless, the concept can be used as well for COTS MPSoCs/ASICS, by applying some modifications. 

\subsection{Code Instrumentation}
\label{subsec:Instrumentation}

\begin{figure}[!t]
\centering
\includegraphics[width=0.45\textwidth]{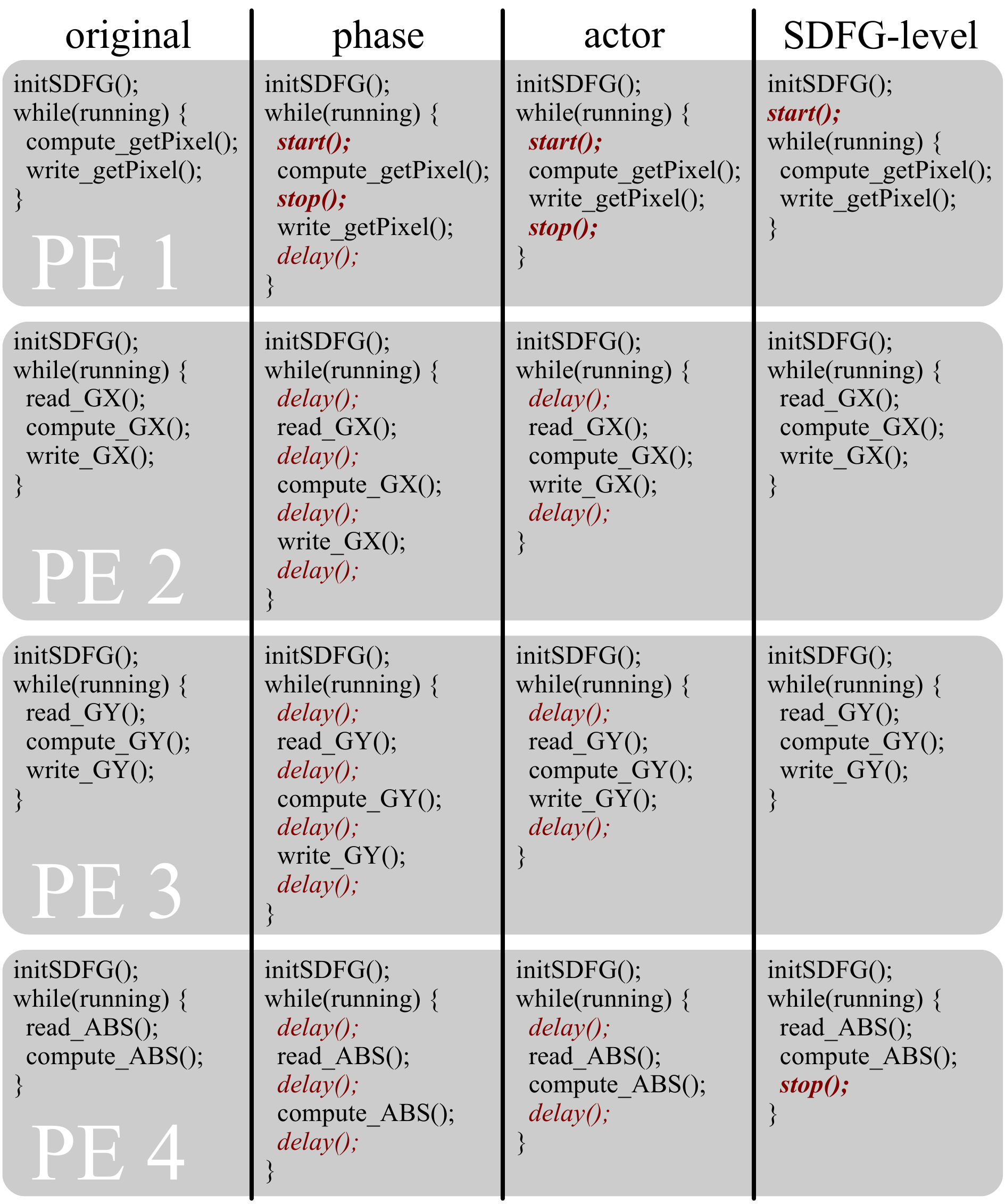}
\caption{Original application code and annotated code on different levels of granularity for a Sobel filter application on a hardware platform with four processing elements (see Fig. \ref{fig_inputs}). As an example for a particular measurement scenario, two delay statements have been replaced by start and stop signals (\textit{bold}) on each granularity level.}
\label{fig_codeannotation}
\end{figure}

When preparing the source code of the application for analysis, the first step is to insert delay statements around related blocks of code, as it can be seen in Fig. \ref{fig_codeannotation}. Depending on the granularity level, the size of these blocks varies. After that, each measurement scenario is created by replacing two consecutive delay statements with start and stop statements. Hence, the code block in between is measured.

For evaluation, we automated this code annotation process with the aid of a script, which transforms an XML-based description of the DUT (including the mapping of the SDFG to the MPSoC) into an instrumented SDF compatible C-code ready to be directly deployed the target processors. This automation significantly speeds up the measurement procedure and reduces errors due to manual implementations.

On the \textit{Phase} granularity level, each code-block that performs a reading, computing or writing operation, is instrumented by delay statements. For the \textit{Actor} granularity level a start measurement statement is put before every actor's read operation and a stop measurement statement is put after their write operations. When annotating on the level of \textit{SDFG} (when tracing the end-to-end latency of an SDFG), we need to start measuring with the execution of first source actor and stop when the last sink actor has completed its computation. Moreover, at the SDFG-level analysis, the measurement must start directly after the previous one is finished, the next iteration follows directly.

In the example depicted in Fig. \ref{fig_inputs} it is easy to detect the beginning and ending of an iteration of the Sobel filter SDFG. Nevertheless, special mechanisms are necessary to deal with SDFGs that contain more than one source or sink actors. The measurement must start, when the first source actor fires. This is achieved by making each processor send a 'start measurement' signal, before it executes a source actor of the considered SDFG.
Whenever the measurement controller receives a start signal, the measurement gets started. Any further start signals are ignored until the measurement is stopped and can restart again.
The end of the measurement is recognized by stop signals. Processors send a stop signal when they finish the execution of their last sink actor of the considered SDFG. The measurement controller counts these incoming signals and stops the measurement only when every processor that fires some sink actors has indicated the end of the sink actors' computation. Due to the fact, that the number of required stop signals varies from an SDFG to another, this parameter must be configured in the measurement controller before the measurement begins. For this purpose, we implemented a software API for configuring the measurement controller (auto-restart, number of stops etc., number of measurements etc.). Since the configuration is done before the actual measurement starts, it does not affect the measurement results (e.g. delay).

If multiple levels of granularity are required for a detailed analysis of an SDF application, the code must be annotated on the lowest level (phase granularity).
It is always possible to annotate the code on phase granularity to have the option to switch the granularity without affecting previous experiments. However, this configuration has the highest impact on the (timing) behavior.

\subsection{Implementation}

\begin{figure}
\centering
\includegraphics[width=0.45\textwidth]{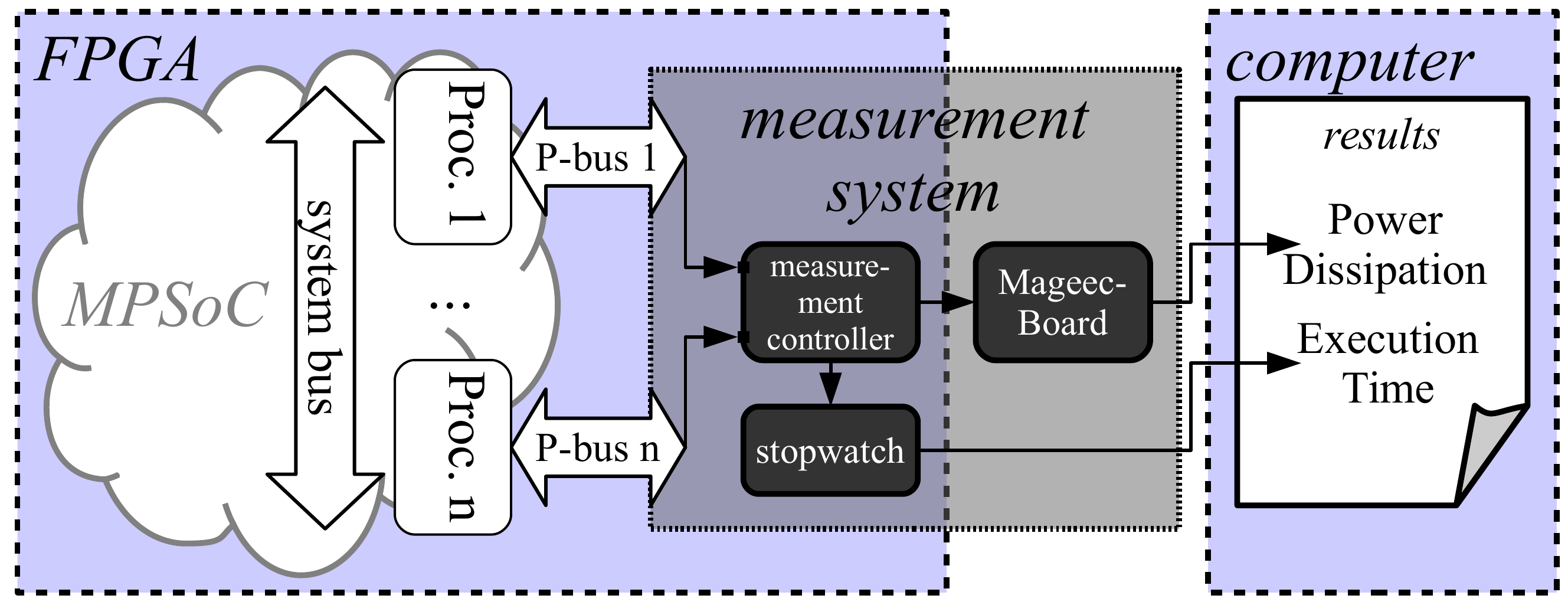}
\caption{Measurement-based approach implemented on an FPGA connected to the \textit{Mageec} power measuring board. The timing is measured on the FPGA, whereas the power is analyzed with the external Mageec-Board. Both parts send their results to the measurement host computer.}
\label{fig_structure}
\end{figure}

The measurement of timing and power consumption is handled in different ways (as shown in Fig. \ref{fig_structure}).
The timing is measured by a 'stopwatch' module, which is included in the FPGA design. 
Since the same clock drives the MPSoC and the stopwatch, we can achieve a cycle-accurate time measurement.

After a timing measurement is performed, the result is stored in a buffer on the FPGA. Once enough values have been obtained or the buffer is full, the measured values are transmitted (via UART) to the computer for further analysis. Depending on how many iterations should be covered in the measurements, the buffer size needs to be properly chosen. Because of this procedure, the power measurement is not influenced by timing value transmission.

For the power measurement, we chose the external Mageec-board, because FPGA-integrated power measurement devices could not cope with our high sample rate demands for measuring short phases of the application. With a sample rate of $84kHz$ the Mageec-board is qualified for this task. However, higher sample rates are still desirable. After customizing the device's firmware by installing a buffer for the measured values, the Mageec-board is able to measure quick successions of short code blocks.

\begin{figure}
\centering
\includegraphics[width=0.45\textwidth]{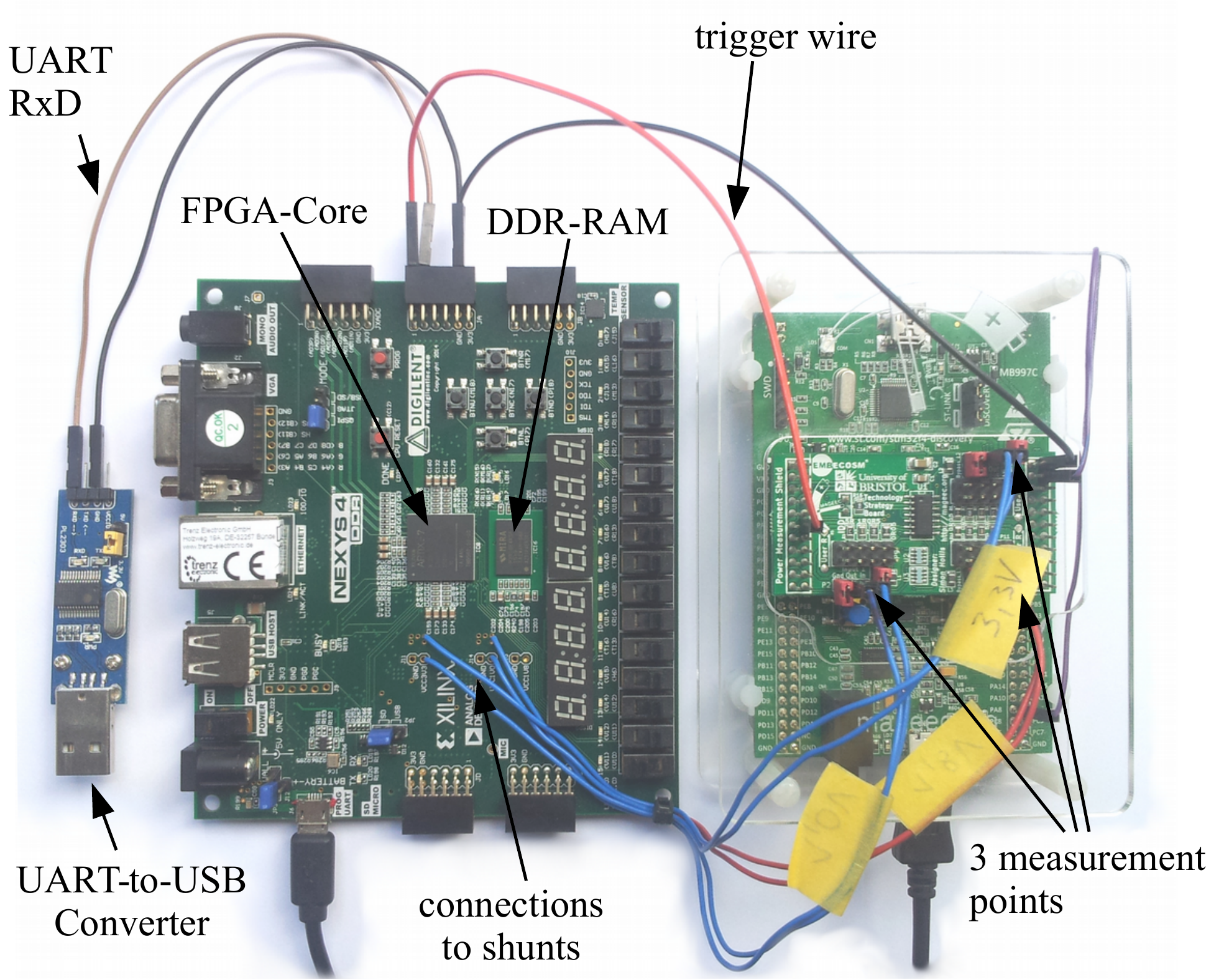}
\caption{Measurement system setup, consisting of an FPGA (Digilent Nexys 4 DDR), and UART-to-USB converter for the transmission of the timing values and the Mageec-board for power measurement.}
\label{fig_setup}
\end{figure}

By applying the shunt resistor method for electrical current measurement \cite{reda_power_2012}, the Mageec-board is able to sample the power consumption of three devices. In our setup with the Digilent Nexys 4 DDR (which is shown in Fig. \ref{fig_setup}), we removed three 'dummy' ($0 \Omega$) resistors (R254, R246 and R261) from the (on-board) wiring of the FPGA-Core, FPGA-IO and the DDR-RAM. After that, we soldered three shunt resistors ($10m \Omega$, $20m \Omega$ and $20m \Omega$) into these places, which values were most appropriate regarding the right balance between high resolution (high values) and stable operation of the FPGA (low values; low voltage reduction).
Both ends of these resistors are connected to the Mageec-board, providing the ability to measure the voltage at each resistor and thus calculating the power consumption.

\section{Evaluation}
\label{sec:Evaluation}

The timing measurement is cycle accurate by design. For verifying this claim, we executed and measured some code blocks with the Xilinx AXI timer. After that, we compared these results with our own timing measurement approach and validated their equality.


Evaluating the accuracy of the power measurement is more challenging. On the one hand, the connection between the measurement controller and the Mageec-board shows some delay. On the other hand, the measured power values include some errors, because of imperfect manufacturing processes of the integrated components (i.e. shunt resistors in the FPGA; amplifiers and ADCs of the Mageec-board) typically having some tolerance intervals (e.g. $\pm 5$ LSB (Least Significant Bit) for the ADC). With the help of the corresponding data-sheets the latter amounts to $\approx 5\%$. Each signal, when sent by the measurement controller and received by the Mageec-board, has a delay 25 cycles in the worst-case. Hence, the beginning and the ending point of the power measurement may be 25 cycles (referring to a $100MHz$ design) late and may include 25 cycles of the following (not ought to be measured) code block.

Due to the sample rate of the Mageec-board, which is lower than the $100MHz$ clock of the MPSoC, the shortest code block noticeable for power measurement must be at least $1200$ cycles (on a $100MHz$ system). The Mageec-board is not capable of measuring shorter code blocks, which is the case in some of the measurements taken in table \ref{tab_results_Sobel} (see 'n/a').

\begin{figure}
\centering
\includegraphics[width=0.45\textwidth]{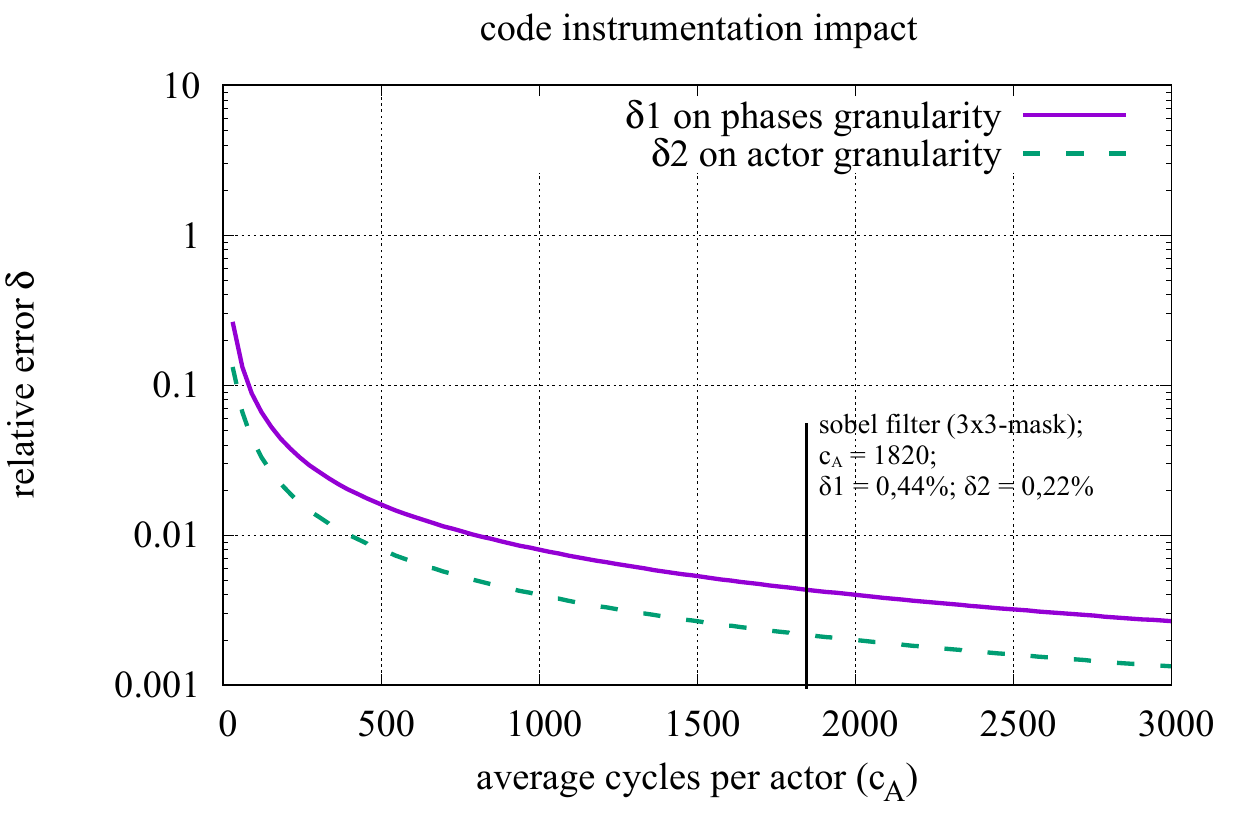}
\caption{Invasiveness impact due to code instrumentation with delay statements. The higher the complexity (execution cycles per actor), the lower the percentage impact. The Sobel filter with a 3x3-mask (which is a rather simple application) needs an average of 1820 cycles per execution of an actor, which is increased by 0.44\% when the code is annotated on phases granularity ($\delta 1$) and 0.22\%, when it is annotated on the actor-level granularity ($\delta 2$).}
\label{fig_invasiveness_impact}
\end{figure}

As mentioned in the concept description, the code annotation will permanently be present in the application, even after the analysis. The comparison between original and annotated code shows, that the delay depends on the measurement granularity and the complexity of the actors in the SDFGs (as shown in Fig. \ref{fig_invasiveness_impact}). If we analyze a very simple Sobel filter application with a 3x3 mask on phases granularity, 
the code annotation increases each actor's execution time by $0.44\%$. Considering a more complex Sobel filter with a 9x9 mask the annotation on phases granularity increases the execution time by less than $0.1\%$.

For demonstrating the benefits of our approach, we constructed two experiments. Both of them employ the hardware platform shown in Fig. \ref{fig_inputs}, realized with the following Xilinx modules: MicroBlazes processors, AXI4 interconnect (as system bus), AXI4 streaming interconnect (as peripheral bus that is connected the measurement controller on the FPGA), AXI BRAM controller and block memory generator (for the shared memory).

In the first experiment, a full detailed analysis of a Sobel filter application on a quad-core MPSoC with the fixed mapping as in Fig. \ref{fig_inputs} was performed. Each scenario takes around 1 minute to be measured. The entire measurement procedure took around 40 minutes, including the instrumentation of the source code and the configuration of the software projects in Xilinx SDK. Table \ref{tab_results_Sobel} shows the results. Contrary to our expectations, we notice that the actors GX and GY take more than nine times longer to read 9 tokens than to write 1 token. The reason for this may be the bus arbitration and polling wait times for these actors. In general, the computation times are rather small compared to the communication (read and write) times.

\begin{table}
\centering
\caption{Analysis of a Sobel filter with a 9x9 mask on a quad-core MPSoC on phases granularity.}
\begin{tabular}{r@{\hspace{0.2em}}l@{\hspace{0.1em}}||r@{\hspace{0.2em}}r@{\hspace{0.2em}}r@{\hspace{0.2em}}|r@{\hspace{0.4em}}r@{\hspace{0.4em}}r@{\hspace{0.4em}}}
\multicolumn{2}{c||}{\textbf{phase}} & \multicolumn{3}{c|}{\textbf{exec. time [cyc.]}} & \multicolumn{3}{c}{\textbf{power [W]}} \\ 
               &               & best  & avg.    & worst & best     & avg.     & worst    \\ \hline \hline
\texttt{getP.} & comp.         & 7875  & 7948,5  & 8055  & $0.5536$ & $0.5978$ & $0.6347$ \\
               & write         & 15079 & 15274.9 & 15460 & $0.5415$ & $0.5643$ & $0.5854$ \\ \hline
\texttt{GX}    & read          & 17664 & 18356.9 & 22582 & $0.5407$ & $0.5596$ & $0.5790$ \\
               & comp.         & 4575  & 4575.0  & 4575  & $0.5403$ & $0.6162$ & $0.6787$ \\
               & write         & 282   & 285,1   & 299   & n/a      & n/a      & n/a      \\ \hline
\texttt{GY}    & read          & 17664 & 18390.1 & 29939 & $0.5419$ & $0.5611$ & $0.5798$ \\
               & comp.         & 4575  & 4575,0  & 4575  & $0.5387$ & $0.6135$ & $0.6780$ \\
               & write         & 282   & 285,2   & 294   & n/a      & n/a      & n/a      \\ \hline
\texttt{ABS}   & read          & 20126 & 23174.7 & 34855 & $0.5418$ & $0.5556$ & $0.5754$ \\
               & comp.         & 52    & 52.0    & 52    & n/a      & n/a      & n/a      \\
\end{tabular}
\label{tab_results_Sobel}
\end{table}

The second experiment uses the same hardware platform but explores different mappings of two SDFGs (see table \ref{tab_multi_sdfg_mappings}): a Sobel filter with a 9x9 mask (actors \texttt{getPixel}, \texttt{GX}, \texttt{GY} and \texttt{ABS}) and a JPEG encoder (see SDFG in \cite{shabbir_ca-mpsoc:_2010}) (actors \texttt{getMB}, \texttt{CC}, \texttt{DCT} and \texttt{VLC}). The actors on every processor are scheduled in static order. However, dynamic scheduling is also supported by our approach but was not tested. All the channels (FIFO-buffer) used for communication between actors were put into the shared memory of the MPSoC to invoke contention on the shared bus.

\begin{figure*}
\centering
\includegraphics[width=\textwidth]{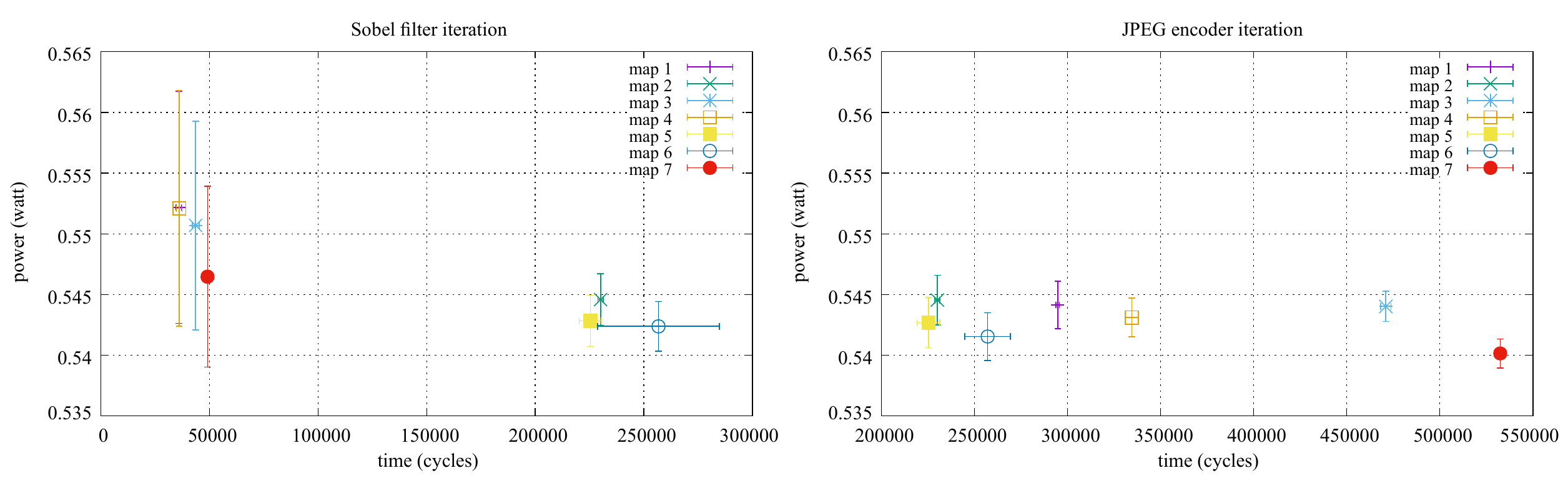}
\caption{Pareto charts for the iterations of the Sobel filter (\textit{left}) and JPEG encoder (\textit{right}) executed on a quad-core MPSoC. Timing and power results vary due to 7 alternative mappings of the actors on the core.}
\label{fig_pareto}
\end{figure*}

The results of the measurements on SDFG-level granularity under different mappings are shown in the Pareto chart in Fig. \ref{fig_pareto}. We can distinguish two groups of mappings of the Sobel filter in Fig. \ref{fig_pareto}. The first group (seen at the bottom-right of Sobel Pareto in Fig. \ref{fig_pareto}) includes mappings 2, 5, and 6 where  the faster SDFG (Sobel filter) actors have a lower priority on the processing elements or have to wait for the slower SDFG (JPEG encoder) actors to finish, before it can finish its own iteration. In the other mappings the SDFGs do not depend on each other and the Sobel filter can reach better end-to-end latency times around 40000 cycles (mappings 1, 3, 4 and 7). Good timing results can be achieved with mapping 1, where the Sobel filter iteration takes an average of 35963 cycles and the JPEG encoder takes 294906 cycles. However, due to the high processor activity and only few waiting times (for polling), the power usage is high in comparison to the other (slower) mappings. The 7th mapping makes use of only two cores, reducing the power but increasing the cycles needed per iteration.


\begin{table}
\centering
\caption{Different mappings of Sobel filter and JPEG encoder actors on a quad-core MPSoC for evaluation.}
\begin{tabular}{c||c|c|c|c}
\textbf{mapping} & \textbf{tile 1}      & \textbf{tile 2}      & \textbf{tile 3}      & \textbf{tile 4}      \\ \hline \hline

\textbf{map 1}   & \texttt{getPixel}    & \texttt{GY}          & \texttt{getMB}       & \texttt{DCT}         \\
                 & \texttt{GX}          & \texttt{ABS}         & \texttt{CC}          & \texttt{VLC}         \\ \hline
\textbf{map 2}   & \texttt{getPixel}    & \texttt{GX}          & \texttt{GY}          & \texttt{ABS}         \\
                 & \texttt{getMB}       & \texttt{CC}          & \texttt{DCT}         & \texttt{VLC}         \\ \hline
\textbf{map 3}   & \texttt{getPixel}    & \texttt{GX}          & \texttt{getMB}       & \texttt{CC}          \\
                 & \texttt{ABS}         & \texttt{GY}          & \texttt{VLC}         & \texttt{DCT}         \\ \hline
\textbf{map 4}   & \texttt{getPixel}    & \texttt{GX}          & \texttt{getMB}       & \texttt{CC}          \\
                 & \texttt{GY}          & \texttt{ABS}         & \texttt{DCT}         & \texttt{VLC}         \\ \hline
\textbf{map 5}   & \texttt{getPixel}    & \texttt{getMB}       & \texttt{GY}          & \texttt{DCT}         \\
                 & \texttt{CC}          & \texttt{GX}          & \texttt{VLC}         & \texttt{ABS}         \\ \hline
\textbf{map 6}   & \texttt{getPixel}    & \texttt{getMB}       & \texttt{GX}          & \texttt{CC}          \\
                 & \texttt{DCT}         & \texttt{GY}          & \texttt{VLC}         & \texttt{ABS}         \\ \hline
\textbf{map 7}   & \texttt{getMB}       & \texttt{getPixel}    &                      &                      \\
                 & \texttt{CC}          & \texttt{GX}          &                      &                      \\
                 & \texttt{DCT}         & \texttt{GY}          &                      &                      \\
                 & \texttt{VLC}         & \texttt{ABS}         &                      &                      \\
\end{tabular}
\label{tab_multi_sdfg_mappings}
\end{table}

\section{Conclusion}

In this paper, we presented a methodology to measure the timing behavior and the power consumption of synchronous data flow applications mapped to FPGA-based MPSoCs.
We showed the viability of our approach being able to measure the timing of SDFAs at cycle accuracy with the help of a customized measurement controller. 
Because the probes used for triggering the measurement unit is replaced by NOP-operations with equivalent timing impact,
the behavior of the measured software will not change significantly in the usage field.
Our power consumption measurement comes with a very low cost solution and is easily applicable to any kind of hardware platform that allows accessing the power supply interconnects.
In the image-processing use-case shown, we demonstrated the benefits of our approach allowing the construction of a Pareto chart of different mappings of actors to the tiles (of the MPSoC) for obtaining power and timing optimal implementations.
Due to limitations of the measurement device using low cost ADCs, our methods works best with measurements at levels above phases granularity. This could be improved in the future by using alternative measurement devices with higher resolution (such as the one in \cite{Weiland2015}).

\section{Acknowledgments}
This work has been partially supported by the ARAMIS II project (01|S16025J), which is funded by the German Federal Ministry of Research and Education (BMBF),
and by the SAFEPOWER project with funding from the European Union's Horizon 2020 research and innovation programme under grant agreement No 646531.

\bibliographystyle{abbrv}
\bibliography{literature}  
%
%


\end{document}